\documentclass[a4paper,fleqn]{cas-dc}

\usepackage[authoryear,longnamesfirst]{natbib}

\def\tsc#1{\csdef{#1}{\textsc{\lowercase{#1}}\xspace}}
\tsc{WGM}
\tsc{QE}

\begin{document}
\let\WriteBookmarks\relax
\def\floatpagepagefraction{1}
\def\textpagefraction{.001}

\shorttitle{Impact of surface reflectivity on the ultra-fast laser melting of silicon-germanium alloys}    

\shortauthors{Ricciarelli et al.}  

\title [mode = title]{Impact of surface reflectivity on the ultra-fast laser melting of silicon-germanium alloys}  



%

\author[1]{Damiano Ricciarelli}[orcid=0000-0003-4213-2514]
\cormark[1]
\cortext[1]{Damiano Ricciarelli}
\ead{see at J. Mater. Sci. Semicond. Proc.}
\credit{Conceptualization, Data curation, Investigation, Visualization, Validation, Formal analysis, Writing - original draft}
\author[1]{Giovanni Mannino}[orcid=0000-0003-2196-6309]
\credit{Investigation, Methodology}
\author[1]{Ioannis Deretzis}[orcid=0000-0001-7252-1831]
\credit{Investigation, Data curation, Writing – review \& editing}
\author[1]{Gaetano Calogero}[orcid=0000-0003-3610-3231]
\credit{Investigation, Data curation, Writing – review \& editing}
\author[1]{Giuseppe Fisicaro}[orcid=0000-0003-4502-3882]
\credit{Investigation, Data curation, Writing – review \& editing}
\author[2]{Richard Daubriac}[orcid=0000-0001-9641-9719]
\credit{Investigation, Data curation}
\author[2]{Fuccio Cristiano}[orcid=0000-0003-1839-9972]
\credit{Investigation, Methodology}
\author[3]{Remi Demoulin}[orcid=]
\credit{Investigation, Methodology}
\author[4]{Paweł P. Michałowski}[orcid=0000-0002-3299-4092]
\credit{Investigation, Methodology}
\author[5]{Pablo Acosta-Alba}[orcid=0000-0003-1965-8067]
\credit{Investigation, Methodology}
\author[5]{Jean-Michel Hartmann}[orcid=0000-0001-7006-8586]
\credit{Investigation, Methodology}
\author[5]{Sébastien Kerdilès}[orcid=0000-0002-6437-172X]
\credit{Investigation, Methodology}
\author[1]{Antonino {La Magna}}[orcid=0000-0002-4087-5210]
\credit{Investigation, Methodology, Conceptualization, Formal analysis, Funding acquisition, Project administration, Software, Supervision, Writing – review \& editing}
\cormark[1]
\cortext[1]{Antonino {La Magna}}
\ead{see at J. Mater. Sci. Semicond. Proc.}

\affiliation[1]{organization={National Research Council, Institute for Microelectronics and Microsystems (IMM-CNR)},
            addressline={VIII Strada 5}, 
            city={Catania},
            postcode={95121}, 
            country={Italy}}
\affiliation[2]{organization={ The French National Center for Scientific Research, Laboratory for Analysis and Architecture of Systems (LAAS-CNRS) and University of Toulouse},
            addressline={7av. Du Col. Roche}, 
            city={Toulouse},
            postcode={31400}, 
            country={France}}
\affiliation[3]{organization={University of Rouen Normandy, INSA Rouen, CNRS, Group of Physics of Materials},
            city={Rouen},
            postcode={76000}, 
            country={France}}
\affiliation[4]{organization={Łukasiewicz Research Network—Institute of Microelectronics and Photonics (IMiF)},
            addressline={Aleja Lotników 32/46}, 
            city={Warsaw},
            postcode={02-668}, 
            country={Poland}}
\affiliation[5]{organization={French Alternative Energies and Atomic Energy Commission, Electronics and Information Technology Laboratory (CEA-LETI), Minatec Campus},
            city={Grenoble},
            postcode={F-38054}, 
            country={France}}
\begin{abstract}
Ultraviolet nanosecond laser annealing (LA) is a powerful tool where strongly confined heating and melting are desirable. In semiconductor technologies the importance of LA increases with the increasing complexity of the proposed integration schemes. Optimizing the LA process along with the experimental design is challenging, especially when complex 3D nanostructured systems with various shapes and phases are involved. Within this context, reliable simulations of laser melting are required for optimizing the process parameters while reducing the number of experimental tests. This gives rise to a virtual Design of Experiments (DoE). $Si_{1-x}Ge_{x}$ alloys are nowadays used for their compatibility with silicon devices enabling to engineer properties such as strain, carrier mobilities and bandgap. In this work, the laser melting process of relaxed and strained $Si_{1-x}Ge_{x}$  is simulated with a finite element method / phase field approach. Particularly, we calibrated the dielectric functions of the alloy for its crystalline and liquid phase using experimental data. We highlighted the importance of reproducing the exact reflectivity of the interface between air and the material in its different aggregation states, to correctly mimic the process. We indirectly discovered intriguing features on the optical behavior of melt silicon-germanium.
\end{abstract}



\begin{keywords}
Laser Annealing \sep Silicon-Germanium \sep Microelectronics \sep Finite Element Methods
\end{keywords}

\maketitle

\section{Introduction}\label{}
Ultraviolet nanosecond laser annealing (LA) with pulsed power emission (pulse duration below $10^{-6}$ s) can be integrated in thermal processes for micro- and nano-electronics, yielding versatile and powerful solutions in extremely constrained space and time scales.
The heat induced by the laser melts the doped semiconductor substrate. During the subsequent re-crystallization, the dopants move from interstitial sites to substitutional sites, becoming activated, and, further, the rapid solidification of the melt material avoids the formation of disordered or amorphous semiconductors domains.
The use of a small wavelength laser results in a melting of well defined regions at the nanoscale with the advantages of a better control of the involved junctions, avoiding possible damage of neighbooring parts of the device.  
The dopant atoms redistribute uniformly due to the high diffusivity ($10^{-4} \, cm^2/s$ in the liquid phase of Si). Moreover, the non-equilibrium segregation during the fast solidification enhances dopant  incorporation in the lattice.  Thanks to these particular characteristics, laser annealing is nowadays widely used as a post-fabrication annealing technique in microelectronics \cite{Prussin, Baeri_1981, ong2004, hernandez2004, svensson2005, huet2009, pilipovich1975}. \par
Optimal control of the process, occurring in a tiny time window of few ns, depending on the laser pulse duration ($\Delta t$), is a key challenge for the successful application of LA. Due to the specificity of the electromagnetic energy absorption and the ultra-rapid thermal diffusion, LA requires a process design which is unique in microelectronics and is complementary to the device design. This complexity impacts the Design of Experiments (DoE) for the optimization of LA processes. Within this context, reliable LA simulations are required to optimize the process parameters while reducing the number of experimental tests \cite{lamagna2004, lombardo2017, lombardo2019, fisicaro2014}. \par
Silicon germanium alloys, $Si_{1-x}Ge_x$, have attracted much interest for decades, notably in the microelectronics industry. They are nowadays used in many  domains. Indeed, their compatibility with silicon devices enables to engineer properties such as strain, carrier mobilities and bandgap thanks to the higher hole mobility in Ge, smaller band-gap and the relatively small lattice parameter deviation \cite{Manku1993, Jain_1991, Iyer1989, people1986, pearsall1989}.
Silicon germanium alloys present peculiar physical properties, for instance the co-presence of Si and Ge in the lattice structure hampers the phononic transport with a consequent U shape of the thermal conductivity vs alloy fraction coordinate (X) \cite{maycock1967, Wagner_2006, takuma2013}. In analogy to pure silicon and germanium, the alloy crystallizes in a diamond-like structure that features semiconductor properties, and it acts as a metal in the liquid phase, with the occurrence of intermediate covalent and metallic bonding frameworks \cite{ko2002}.  \par
In this work, we performed XeCl excimer laser melting simulations of silicon-germanium substrates employing a finite element/phase field approach and a custom-built developed software. This solves coupled partial differential equations (PDEs) which rule evolving fields during the pulsed irradiation (e.g., electromagnetic field, temperature, phase, alloy fraction, dopant density, etc) \cite{lamagna2004, lombardo2017, lombardo2019, fisicaro2014}. This computational methodology was previously applied for the laser annealing of silicon and silicon-germanium, limited to strained thin samples with 0.2 Ge content. \cite{huet2020} The calibration of material parameters is fundamental to achieve the full description of the laser melting process, particularly, in previous work, a systematic categorization of the physical parameters, required for the successful simulation of LA processes on $Si_{1-x}Ge_x$ , was reported \cite{huet2020}. However, critical issues with respect to the  calibration of the dielectric functions of solid and liquid $Si_{1-x}Ge_x$  were also identified, showing a high level of difficulty due to their possible dependencies on the alloy fraction and dopant concentration. The correctness of dielectric functions is crucial to reproduce the air/$Si_{1-x} Ge_x$ interface reflectivity of the sample, which in turns governs the heat transfer from the laser to the specimen \cite{fisicaro2014}. \par
Here, we calibrated the dielectric function of crystalline $Si_{1-x}Ge_x$ at different temperatures, stoichiometries and dopant concentrations by means of spectroscopic ellipsometry. We further fine-tuned the dielectric function of liquid $Si_{1-x}Ge_x$, with an indirect approach, to reproduce the exact melt depth from laser irradiation of relaxed thick samples.
The final dielectric function expressions achieved for both crystalline and liquid phase, were validated for the laser annealing simulation of strained thin samples with various Ge content. \par
Our results show a reasonable agreement between computed and experimental melt depths, confirming that a correct reproduction of the specimen reflectivity, obtained by considering the dielectric functions dependency on alloy fraction and temperature, is key to realistically model the entire laser melting process. \par
One additional finding from our investigation pertains to the unique reflectivity observed in liquid $Si_{1-x} Ge_x$. We observed that the reflectivity is maximized when the germanium content is at an intermediate level and decreases as the temperature rises. This particular behavior may arise from the metallic-like character of the liquid and it deserves further investigations.

\section{Material and Methods}\label{}

The methodology employed in this paper involves several steps. First, we determined the dielectric functions of both p-doped and undoped crystalline strained $Si_{1-x}Ge_x$ samples. Next, we manufactured relaxed thick $Si_{1-x}Ge_x$ samples and subjected them to laser annealing. Then, we fine-tuned a custom-built software to realistically simulate laser melting. To this aim, we used experimentally measured values of dielectric functions to calibrate the reflectivity of crystalline $Si_{1-x}Ge_x$ on the software, and we used the melt depths of laser annealed relaxed thick samples to indirectly reproduced the air/liquid $Si_{1-x}Ge_x$ interface reflectivity on the software.
Finally, the obtained computational model was tested using previously published datasets of laser annealed strained samples subjected to similar irradiation conditions. \cite{dagault2019, dagault2020_2, huet2020}. Details on the technical of all experimental and computational steps follow in the next paragraphs.

\subsection{\label{sec:level2}Strained Samples Manufacture }
We fabricated a set of strained $Si_{1-x}Ge_x$ undoped $\sim$ 30nm thick films over a Si substrate (obtained from epitaxial growths below the critical thickness) with 0.1, 0.2, 0.3, and 0.4 Ge alloy fraction, and a set of p-doped (with boron) strained $Si_{1-x}Ge_x$ samples with 0.3 Ge content. The Ge profiles have been measured with Secondary Ions Mass Spectroscopy (SIMS) and the dopant concentrations achieved, $C_B$, were:  $ 0 \, cm^{-3}$ (None), $7.3 \cdot 10^{19} \, cm^{-3}$ (Low), $1.4 \cdot 10^{20} \, cm^{-3}$ (Medium) and \, $2.3 \cdot 10^{20} \,  cm^{-3}$ (High). \par

\subsection{\label{sec:level2} Crystalline $Si_{1-x}Ge_x$ Dielectric Functions Determination }
The optical functions of the strained samples mentioned above were quantified by means of spectroscopic ellipsometry. This analysis was performed with a J. A. Woollam VASE ellipsometer. The light source consists of a Xe lamp and a monochromator. Wavelength-by-wavelength measurements were conducted, in equilibrium conditions, at variable temperatures, starting from room temperature, i.e. 298 K, up to 873 K, thanks to an Instec closed chamber with a costant incident flux of $N_2$. The presence of the monochromator was particularly important, as it enabled optimized measurement steps in specific spectral regions (i.e., a wavelength step of 0.5 nm in the range 300-320 nm, corresponding to a photon energy step of 0.007 eV in the range 4.13-3.87 eV) which are relevant for laser melting processes. A wider step was used elsewhere within the range 1-6 eV. \par
The dielectric functions were obtained by fitting the experimental data with a dispersion formula based on Tauc-Lorentz oscillators (see Figure S1). For the purpose of this investigation, we consider only dielectric functions values evaluated at 4.02 eV, corresponding to the wavelength of XeCl excimer laser (308 nm).

\subsection{\label{sec:level2}Relaxed Samples Manufacture }
Relaxed thick samples were prepared from two 200 mm bulk Si(100) wafers (Czochralski, p-type, 1–50 $\Omega$ cm). The whole $Si_{1-x}Ge_x$ layers epitaxy process was performed by reduced pressure chemical vapor deposition (RPCVD) in a Centura 5200C epitaxy chamber. Prior to each  $Si_{1-x}Ge_x$ layer epitaxy, a  $H_2$  bake (1373 K, 2 min) was done to remove the native oxide. After the surface cleaning, a graded $Si_{1-x}Ge_x$ buffer layer was grown on each wafer, with specific growth conditions to reach X=0.2 Ge content for one wafer and X=0.5 Ge content for the other one, with a $10\% / \mu m$ ramp (T(X=0.2) = 1173 K and T(X=0.5) = 1123 K, P = 20 Torr, precursors: $SiH_2Cl_2 + GeH_4$). Then, 1.2 $\mu m$ thick relaxed and undoped $Si_{1-x}Ge_x$ layers were grown with a uniform Ge content of 0.2 and 0.5, corresponding to the Ge content of the buffer layer underneath. Thanks to the high temperature used during the process, the glide of the threading arms of misfit dislocations (i.e. threading dislocations) was enhanced in such way that they remained mostly confined in the graded buffer layers, close to the $ Si_{1-x}Ge_x /Si$ interface. As a result, the threading dislocations density was significantly reduced in the $Si_{1-x}Ge_x$ top layers ($\sim 10^5 cm^2$). Following the RPCVD process, the remaining cross-hatch patterns were removed using a two steps (planarization and smoothing) chemical-mechanical polishing (CMP) process thanks to a Mirra CMP system, reducing the thickness of the $Si_{1-x}Ge_x$ top layers from 1.2 µm to $\sim 0.7 \mu m$. \par

\subsection{\label{sec:level2}Laser Annealing}

Ultraviolet nanosecond laser annealing treatments were performed with a  UV laser of type SCREEN-LASSE (LT-3100) with a $\lambda$ of 308 nm, a pulse duration of 160 ns, 4 Hz repetition rate, $<3 \%$ laser beam uniformity and $10 \cdot 10 mm^2$ laser beam. The samples were kept at room temperature and atmospheric pressure, with a constant incident $N_2$ flux. Single pulse anneals at various energy densities (ED) were carried out, ranging from 0.300 to 2.500 $J cm^{-2}$ with a 0.025 $J cm^{-2}$ incremental step, crossing all main laser regimes, from sub-melt to partial melt of the relaxed $Si_{1-x}Ge_x$  layers. \par
The germanium composition of as-deposited and laser irradiated $ Si_{1-x}Ge_x$ layers was measured with the Energy Dispersive X-ray spectroscopy (EDX) technique implemented in a high-angle annular dark-field scanning transmission electron microscope (STEM-HAADF)  of JEM-ARM200F model. The technique was previously tested and calibrated, see the Supporting Information for more details.

\subsection{Laser Melting Simulations}\label{}
Numerical simulation of the ultrafast laser melting process, involving solid/liquid phase change, and Ge diffusion phenomena were performed employing a pre-existing finite element method / phase field approach and a custom-built developed code \cite{lamagna2004, fisicaro2014, lombardo2017, huet2020}. This tool consists of a Technology Computed-Aided Design (TCAD) package able to simulate the laser annealing process for 1D, 2D and 3D structures \cite{lombardo2019}. The heat equation, coupled to the time-harmonic solution of Maxwell equations is solved self-consistently including phase and temperature dependency of material parameters, phase change and alloy fraction. The core model equations, more detailed in \cite{lamagna2004, fisicaro2014, lombardo2017, huet2020}, are provided in (1)–(4).
\begin{equation}
 S(r,t) = \frac {\Im (\varepsilon)}{2 \rho} | {E_{t-h}} |^2
\end{equation}
\begin{equation}
 \rho c_p \frac{\partial T}{\partial t} + [(1-X) L_S + X \Delta L_{S-X}] \frac{\partial h}{\partial t} = \nabla [K \nabla T] + S(r,t)
\end{equation}
\begin{equation}
 \frac{\partial{\varphi}}{\partial t} = D_{\varphi} \nabla^2(\varphi)-\frac{\partial F (\varphi, g, h, u)}{\partial \varphi}
\end{equation}
\begin{equation}
 \frac{\partial X}{\partial t} = \nabla [D_X \nabla X]-D_X ln(k) \nabla [M_2 X (1-X) \varphi (1 - \varphi) \nabla \varphi]
\end{equation}
where $r$ is the position, $t$ is the time, $\varphi$ is the phase (1 for solid and 0 for liquid), T the temperature, $X$ the solute species molar fraction and $u$ the normalized enthalpy respectively. $F$ is the Helmoltz free energy functional, $f$ the Helmoltz free energy density, $g$ and $h$ are related functions chosen to obtain the adequate shape of $F$ in the sharp interface limit. $S(r,t)$ is the heat source due to the laser, $\Im (\varepsilon)$ is the imaginary part of the dielectric function of the material and $E_{th}$ is the time harmonic electric field (from the solution of the corresponding Maxwell equations).  $L_S$ is the latent heat of the pure material, $\Delta L_{S-X}$  the latent heat change due to the solute, $\rho$ the density,$c_p$  the specific heat, $K$ the thermal conductivity, $k$ the equilibrium segregation coefficient, $D_X$ the solute species diffusivity and $M_2$ the solute species mobility coefficient. \par
The interface velocity is  evaluated with the following expression from \cite{mittiga2000}:
\begin{equation}
v(T)=A \cdot exp { \Bigl( {-\frac{E_a}{k_bT} \Bigr) } } \cdot \Bigl[ 1-exp \Bigl[ \Bigl( \frac{\rho L}{k_B N} \Bigr) \Bigl( \frac{1}{T_M}-\frac{1}{T} \Bigr) \Bigr] \Bigr]
\end{equation}
where $A$ is the velocity prefactor, $E_a$ is the activation energy for the transition of the atoms from the liquid to the solid phase, $k_B$ the Boltzmann constant, $N$ the atomic density, $T_M$ the melting temperature of the material and $L$ is the latent heat.\\
As shown by equations (1)-(4) the evolving physical fields are the temperature, $T$, alloy fraction, $X$ and phase field, $\varphi$, they clearly depend on space and time and their initial value (t = 0 ns) must be initialized by the user (for all the points of the mesh).  \\
To save computational resources, the phase field model (defined by the above equations) is activated only when $T>T_{melt}$ for a spatial region greater than a user-defined threshold, normally $\sim$ 8 nm in the mono-dimensional case, while the less costly enthalpy model is used for the other cases \cite{LOMBARDO2021251}.
The input CAD geometries and their mesh are built using the gmsh software \cite{gmesh}. The partial derivative equations are solved self-consistently by the FENICS computing platform \cite{fenics2015}. \par
With the exception of solid and liquid $Si_{1-x}Ge_x$ dielectric functions, calibrated in this work, and of the thermal conductivity, taken from \cite{Wagner_2006}, the functions of the material properties involved in the partial derivative equations can be found in \cite{huet2020}, where the same custom-built developed software was used.

\section{Results and Discussion}\label{}
The laser melting process, as introduced, is sensitive to the fraction of radiation absorbed by the material, i.e the photons not reflected by the surface of the sample. In Figure 1, we show a simplified scheme of the process. The radiation is partly absorbed by the first few $Si_{1-x}Ge_x$ layers and the electromagnetic energy becomes thermal energy. When the alloy reaches the melting temperature, i.e. when the energy density of the laser overcomes a certain ED threshold, the first liquid nuclei start to form. Then, for the duration of the laser pulse $\Delta t$, the liquid front covers a distance called melt depth and, eventually, the sample re-crystallizes. \par
ED thresholds and melt depths are related to the fraction of radiation absorbed, $(1-R)$, with $R$ being the surface reflectivity. The reflectivity in object is related to the air/$Si_{1-x}Ge_x$  interface, as the absorption of the electromagnetic radiation extinguishes within the first nm of the material, i.e. $\sim$ 5 nm. To accurately simulate laser melting is important to consider the dielectric functions of $Si_{1-x}Ge_x$ for crystalline and liquid phases. The first one will mainly determine the melt threshold of the sample, while the second is crucial to obtain the exact melt depth.  The link between reflectivity and dielectric function of a certain system at the interface with air is provided by real and complex refractive indices, $n$ and $k$, through expressions (6)-(8). \par
\begin{equation}
 n=\sqrt{{\Bigl[\Re (\varepsilon)+\sqrt{(\Re\ {(\varepsilon))}^2+(\Im\ {(\varepsilon))}^2}}\Bigr]/{2}}%
\end{equation}
\begin{equation}
 k={\Im (\varepsilon)}/{(2n)}\ \ \ \ 
 \end{equation}
\begin{equation}
 R={[\left(n-1\right)^2+k^2]}/{[\left(n+1\right)^2+k^2]}%
\end{equation}
\begin{figure}
\includegraphics[scale=0.75]{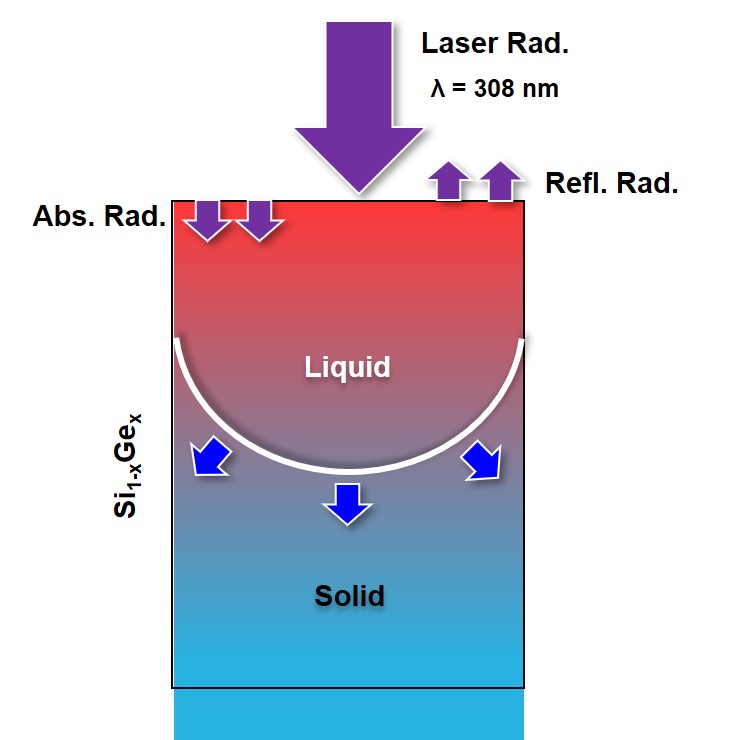}
\caption{\label{figure:wide} Schematics of the laser-melting process with laser radiation (Laser Rad.), absorbed radiation (Abs. Rad.) and reflected radiation (Refl. Rad.). Warmer areas are red-coloured , colder ones blue-coloured.}
\end{figure}
We started to determine the dielectric function of crystalline $Si_{1-x}Ge_x$, $\varepsilon_c$, performing a three-dimensional fitting of $\Re (\varepsilon_c)$ and $\Im (\varepsilon_c)$ against temperature (T) and alloy fraction (X). To that end we performed spectroscopic ellipsometry measurements on strained samples, detailed in the methodological section. We considered quantified $\Re (\varepsilon_c)$ and $\Im (\varepsilon_c)$ values  corresponding to the XeCl excimer laser wavelength of 308 nm, i.e. 4.02 eV, see Figure S1 for more details on the fitting procedure. The studied alloy fractions were 0.1, 0.2, 0.3 and 0.4, while the temperature range spanned from 295K to 853K. \par
We first inspected the experimental data reported in Figure 2a as a function of T. $\Re (\varepsilon_c)$ monotonously decreased while $\Im (\varepsilon_c)$ increased with the temperature. This, as shown in the lower part of the panel, resulted in a progressively increased reflectivity with temperature. On the other hand, alloy fraction variations, from 0.1 to 0.4, mainly impacted the imaginary parts, as shown in Figure 2a, with $\Im (\varepsilon_c (0.1)) > \Im (\varepsilon_c (0.2)) > \Im (\varepsilon_c (0.3)) > \Im (\varepsilon_c (0.4))$. This, turning to reflectivity graphs (lower part of the panel), resulted in a slight decrease of R with the sample alloying. \\
The effect of p-doping was further assessed for a fixed X=0.3 alloy fraction. Three boron  concentrations, $C_B$, were evaluated: low $7.3 \cdot 10^{19} \, cm^{-3}$, medium $1.4 \cdot 10^{20} \, cm^{-3}$ and high $2.3 \cdot 10^{20} \, cm^{-3}$.
The real and imaginary parts of dielectric function at 308 nm, shown in Figure 2b, presented similar variations in sign and amount, with  $\varepsilon_c (low) > \varepsilon_c (medium) > \varepsilon_c (high)$. In terms of reflectivity we found that this results in slight variations among the different doping concentrations, as shown in the bottom of the panel. \par
\begin{figure*}
\begin{center}
\includegraphics[scale=0.50]{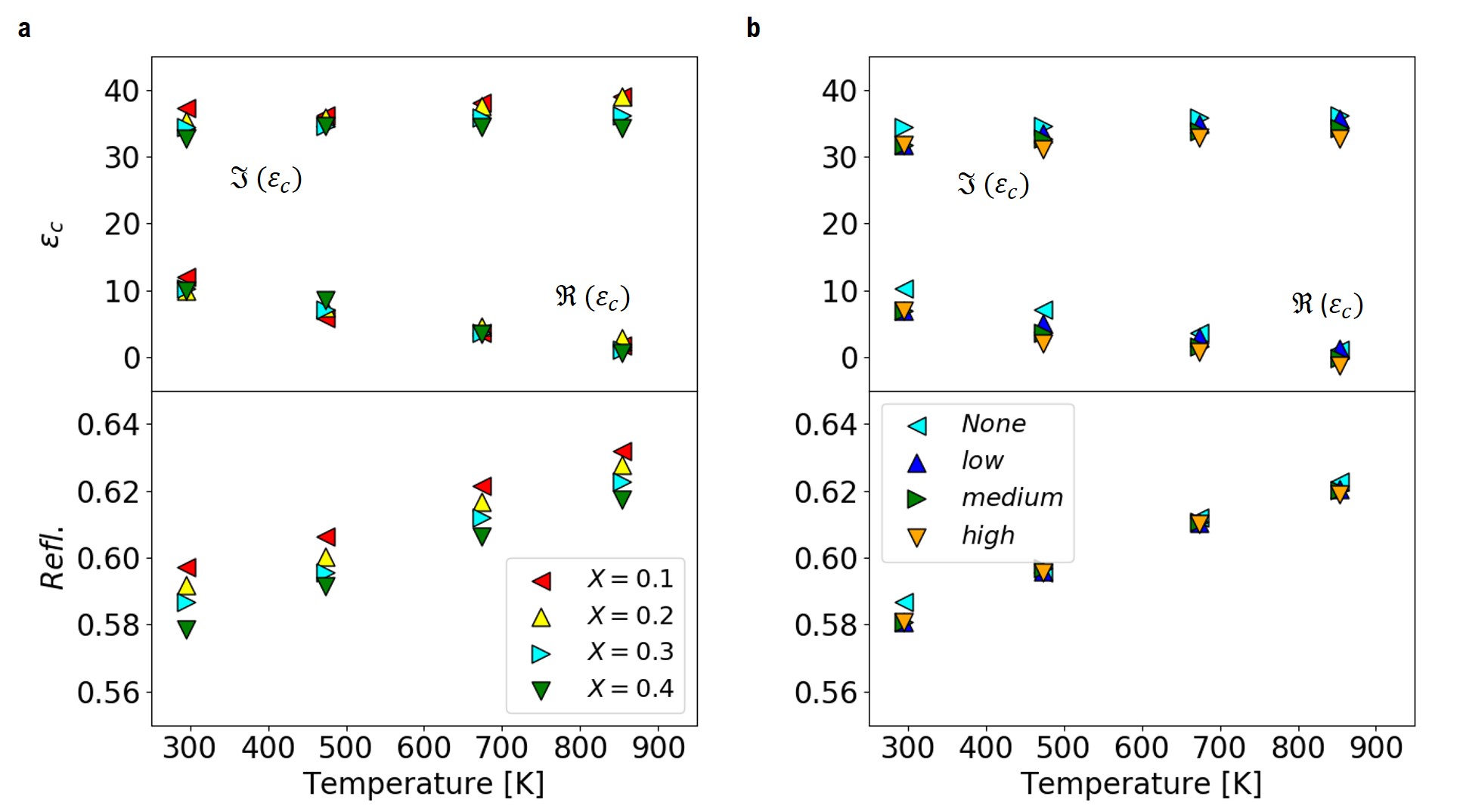}
\caption{\label{figure:wide}  Imaginary, $\Im (\varepsilon_c)$, real, $\Re (\varepsilon_c)$ parts of dielectric functions values and associated reflectivities measured at 308 nm by spectroscopic ellipsometry at different temperatures and alloy fractions for undoped $Si_{1-x}Ge_x$ (a), and X=0.3 boron doped $Si_{1-x}Ge_x$  samples (b).}
\end{center}
\end{figure*}
\begin{figure*}
\begin{center}
\includegraphics[scale=0.65]{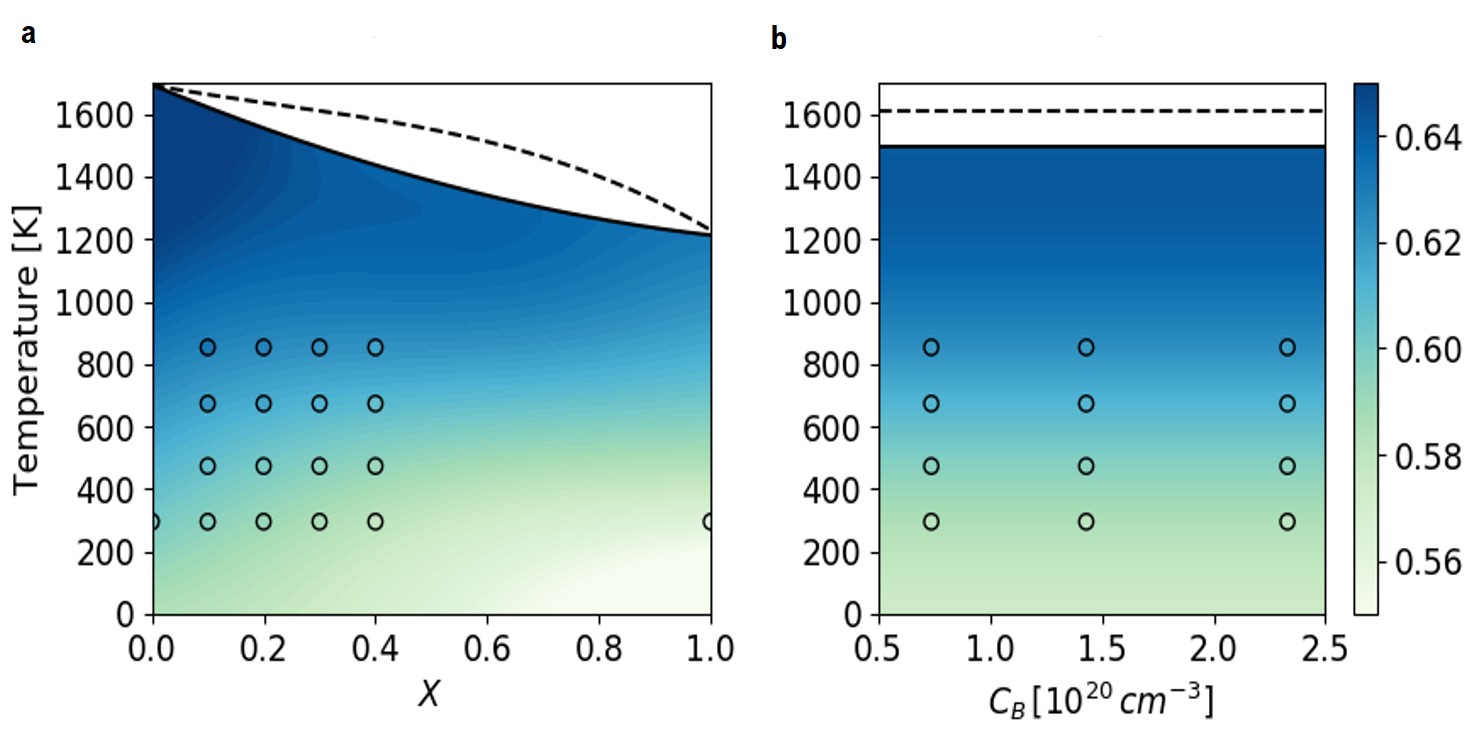}
\caption{\label{figure:wide}  Reflectivity of $Si_{1-x}Ge_x$ obtained from fitted $\varepsilon_c$ for undoped (a) an p-doped X=0.3 samples (b). The continuous line represents the solidus thermodynamic limit and the dashed one the liquidus, redrawn from \cite{olesinski1984}. Coloured dots represent the experimental values.}
\end{center}
\end{figure*}
\begin{table*}
\caption{\label{tab:table3} List of fitting parameters for the real, $\Re$, and imaginary $\Im$ parts of the dielectric function for doped and undoped crystalline $Si_{1-x}Ge_x$. For definitions cf. main text. Temperature is expressed in K.}
\resizebox{\textwidth}{!}{\begin{tabular}{ccc}
\toprule
&$\Re$&$\Im$\\
\midrule
$\varepsilon_{c,{Si}}$&$(3.912 \cdot 10^{-6}) \cdot T^2 -(1.355 \cdot 10^{-2}) \cdot T +8.941$  &$(-5.225 \cdot 10^{-6}) \cdot T^2 +(1.593 \cdot 10^{-2}) \cdot T +23.571$\\
$\varepsilon_{c,{Ge}}$&$(-9.025 \cdot 10^{-7}) \cdot T^2 -(6.558 \cdot 10^{-3}) \cdot T +13.892$&$(9.652 \cdot 10^{-3}) \cdot T+35.069$\\
$b$&$2.670 \cdot 10^{-6}$&$6.167 \cdot 10^{-7}$\\
$c$&$-8.932 \cdot 10^{-3}$&$-3.678\cdot 10^{-3}$\\
$d$&$2.747$&$0.7241$\\
$b'$&$-3.043 \cdot 10^{-8}$&$-1.555 \cdot 10^{-9}$\\
$c'$&$8.797 \cdot 10^{-5}$&$5.517 \cdot 10^{-7}$\\
$d'$&$-4.731 \cdot 10^{-3}$&$4.131 \cdot 10^{-3}$\\
$C_0$&$1.00 1 \cdot 10^{19}$ &$1.00 \cdot 10^{19}$\\
\bottomrule
\end{tabular}}
\end{table*}

\begin{figure*}
\includegraphics[scale=0.5]{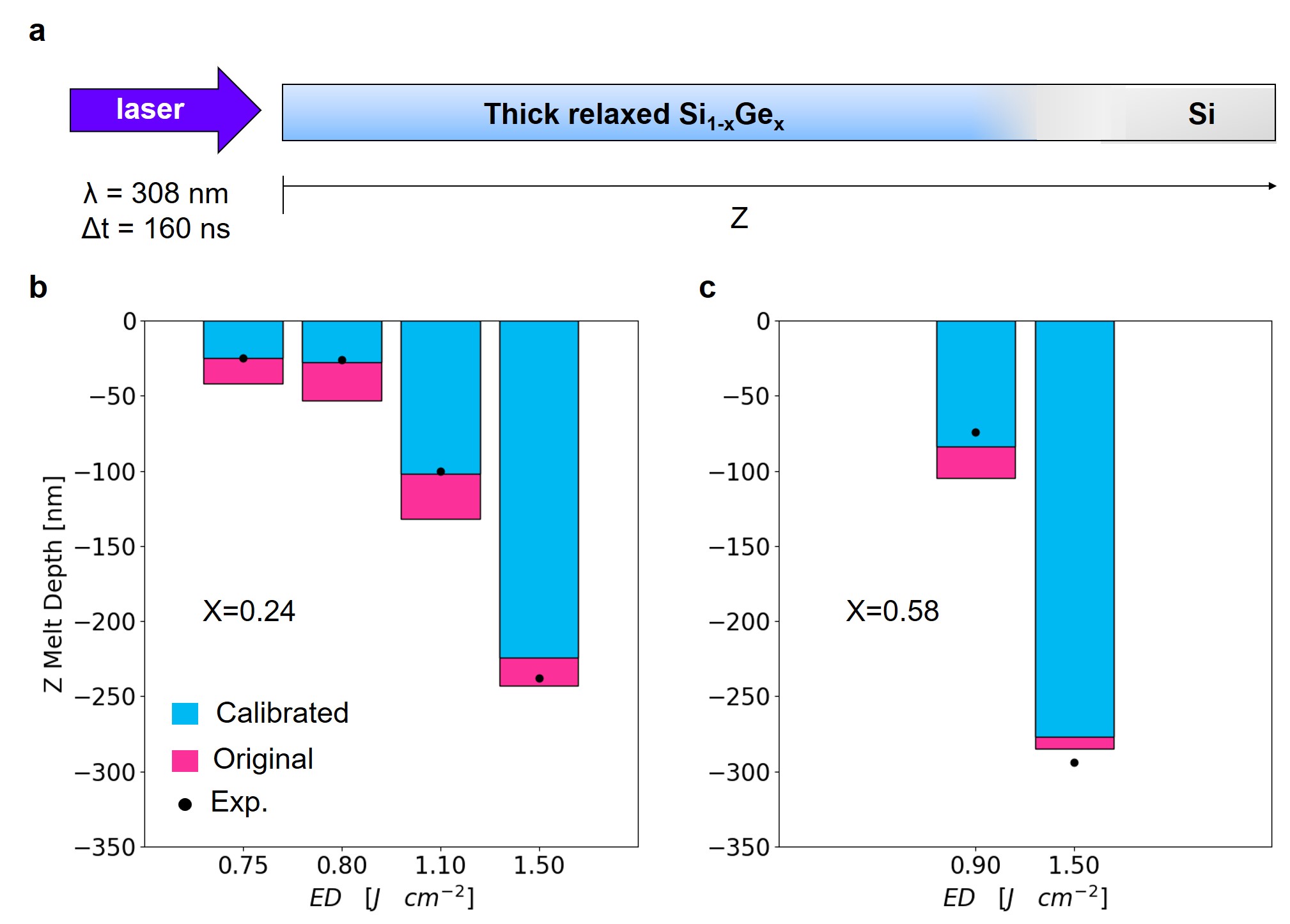}
\caption{\label{figure:wide} Schematics of the mono-dimensional model employed for simulating laser annealing on relaxed $Si_{1-x}Ge_x$ samples (a), comparison between experimental and simulated MD for relaxed $Si_{1-x}Ge_x$ samples with initial alloy fractions of 0.24 (b) and 0.58 (c).  Cf. main text for definition of calibrated and original models }
\end{figure*}
 The fitting performed is structured on the basis of the following theoretical scheme: $Si_{1-x}Ge_x$ is an almost ideal binary alloy system, where Si and Ge are fully miscible over the whole range of composition. This generally makes the linear interpolation between the physical properties of Si and Ge (using the Ge alloy fraction variable X) a good starting point for the calibration of this material. However, some critical uncertainties exist. A more accurate determination of the dependence of the optical parameters on X in each phase is necessary. We expressed the real and imaginary parts of the optical dielectric function $\varepsilon_c$ as: \\
\begin{equation}
 \varepsilon_c(T,X)  = \varepsilon_{c , Ge} (T) \cdot f(X,T) + \varepsilon_{c , Si}  \cdot [1 - f(X,T)]
\end{equation}
where $f(X,T)$ is a monotonically growing polynomial function satisfying the relationships $f(0,T) = 0$ and $f(1,T)=1$, while $\varepsilon_{c \, Ge} (T)$ and $\varepsilon_{c \, Si} (T)$ are the Ge and Si functions reported in Table 1 from \cite{huet2020}. Only $f(X,T)$ has an unknown form and calibration. We thus considered a second order polynomial function:
\begin{equation}
f(X,T) = a(T) \cdot X^2 + [1 - a(T)] \cdot X\\
\end{equation}
In order to determine the temperature dependence of $\varepsilon_c$, the function $a(T)$ was further calibrated as a second-order polynomial:
\begin{equation}
a(T) = b \cdot T^2 + c \cdot T + d\\
\end{equation}
A second level of calibration was implemented for p-doped $Si_{1-x}Ge_x$ samples. In this case, the variable $C_B$ was introduced in the $\varepsilon_c$ function giving rise to $\varepsilon_c(T,X,C_B)$:
\begin{eqnarray}
\varepsilon_c(T,X,C_B) = \varepsilon_c(T,X) \cdot g(C_B,T)
\\
g(C_B,T) = 1 - m(T) \cdot {C_B}/{C_0}
\\
m(T) = b’ \cdot T^2 + c’ \cdot T + d’
\label{eq:one}.
\end{eqnarray}
$m(T)$ is a second-order polynomial function of temperature with parameters $b’$, $c’$ and $d’$, while $C_0$ is a constant yielding $g(C_B,T) \approx 1$ for very low-doping (hence, for very low doping, $\varepsilon_s(T,X,C_B) \approx \varepsilon_s(T,X)$. All fitting parameters were reported in Table 1. We note that the expressions for crystalline Si and Ge were obtained from parameterizations achieved in Ref \cite{huet2020} by some of us. \\
Figure 3a-b shows the calculated reflectivity map (from our fitting expression) for undoped (3a) and boron-doped (Figure 3b) samples. Coloured dots representing the experimental values were also plotted showing on overall a good agreement between the mathematical model and spectroscopic ellipsometry data. Numerically the errors associated to the modeled reflectivity result within $\sim 5 \%$, as detailed in Table S1 and S2. \\
A closer look to Figure 3a highlights a huge dependency of crystal R on the temperature, governed by the material phonons, with a remarkable steepness, while only slight variation can be found when moving along X.\\ Importantly, we observed that, in the large  $T$ and $X$ ranges, the crystalline $Si_{1-x}Ge_x$ reflectivity has an average value of $\sim$ 0.60.\\
In the case of p-doped material, Figure 3b, we observed slight dependence of R on the dopant concentration.\\
\begin{table}[b]
\caption{\label{tab:table4} Original vs optimal $\Im (\varepsilon_l)$ values obtained for matching the melt depth at the various laser energy densities. $R_l$ represents liquid reflectivity, $T_m$ the melting point of the alloy and $T_l$ the liquid temperature collected at the interface with air (cf. main text).}
\resizebox{\columnwidth}{!}{\begin{tabular}{|cc|cccc|ccc|}
\toprule
\multicolumn{2}{|c|}{}&\multicolumn{4}{c}{Original $\Im \, (\varepsilon_l)$}&\multicolumn{3}{|c|}{Optimal $\Im \, (\varepsilon_l)$}\\
\midrule
$X_l$&ED&$\Im\, (\varepsilon_l)$&$R_l$&$T_m$&$T_l$&$\Im\, (\varepsilon_l)$&$R_l$&$T_l$\\
&$[J/cm^2]$&&&[K]&[K]&&&[K]\\
&&&&&&&&\\
0.24&0.75&9.98&0.778&1626&1641&7.70&0.811&1640\\
0.24&0.80&9.98&0.778&1626&1643&6.50&0.833&1641\\
0.24&1.10&9.98&0.778&1626&1660&8.01&0.806&1652\\
0.24&1.50&9.98&0.778&1626&1696&9.80&0.780&1693\\
0.58&0.90&9.77&0.780&1520&1562&7.90&0.803&1557\\
0.58&1.50&9.77&0.780&1520&1627&10.30&0.768&1630\\
\bottomrule
\end{tabular}}
\end{table}
\begin{table}[b]
\caption{\label{tab:table4} Calibrated parameters for the real and imaginary dielectric functions of liquid $Si_{1-x}Ge_x$. For definitions cf. main text. Temperature is expressed in K.}
\setlength{\tabcolsep}{25pt}
\resizebox{\columnwidth}{!}{\begin{tabular}{ccc}
\toprule
&$\Re$&$\Im$\\
\midrule
$\varepsilon_{l,Si}$&-15.734&10.126\\
$\varepsilon_{l,Ge}$&-14.585&9.517\\
$b_1$&-&0.8787\\
$c_1$&-&-496.7\\
$b_2$&-&-0.4159\\
$c_2$&-&257.9\\
\bottomrule
\end{tabular}}
\end{table}
\begin{figure}
\begin{center}
\includegraphics[scale=0.40]{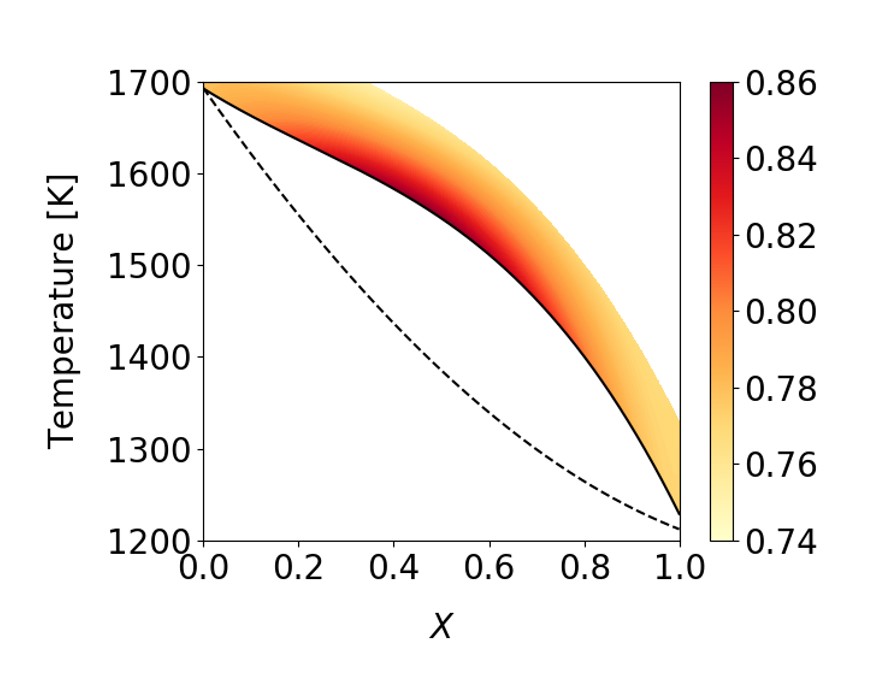}
\caption{\label{fig:wide}Reflectivity map of liquid $Si_{1-x}Ge_x$ at a wavelength of 308 nm as a function of temperature and alloy fraction. The continuous line represents the liquidus and the dashed one the solidus thermodynamic limits, redrawn from \cite{olesinski1984}.}
\end{center}
\end{figure}
\begin{figure*}
\includegraphics[scale=0.425]{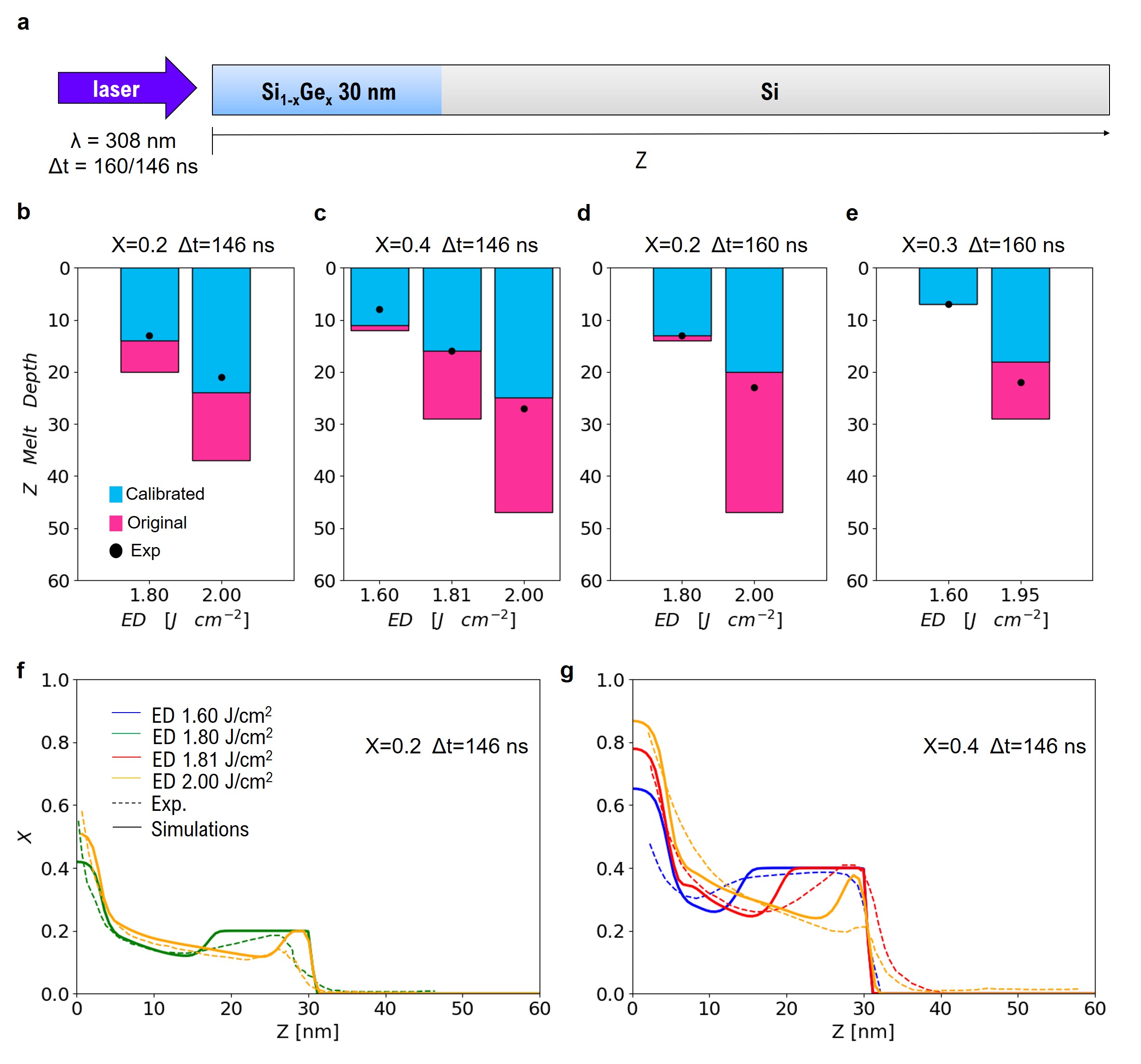}
\caption{\label{figure:wide} Schematics of the mono-dimensional model employed for laser annealing simulations on strained $Si_{1-x}Ge_x$ samples (a), comparison between experimental and simulated MDs for a laser pulse of 146 ns (b)(c) and 160 ns (d)(e), experimental vs simulated germanium profile for X=0.2 (f) and X=0.4 (g) with $\Delta t$ of 146 ns, the experimental profiles are re-drawn from \cite{dagault2019, dagault2020_2}. Cf. main text for definitions of original and calibrated models.}
\end{figure*}
\par
Having obtained a reasonable calibration for the real and imaginary dielectric functions of crystalline $Si_{1-x}Ge_x$, we can now start to use our software to model the laser melting process, however results need to be compared to the experiment. Experimental laser irradiations were performed on relaxed thick samples.  The use of relaxed thick samples enabled to evaluate cases where the liquid front covered high distances.  A XeCl excimer laser was used, with a wavelength of 308 nm and a laser pulse $\Delta t$ of 160 ns. Irradiated samples preserved an optimal surface planarity, as shown by Figure S3. The samples exhibited a constant X alloy fraction of 0.24 and 0.58 for $\sim 1200 \, nm$ and $\sim 3000 \, nm$ respectively evaluated by EDX measurements (see Figure S2a-b). These numbers quite differed from the nominal alloy fraction and thickness of the samples, i.e. 0.20 (0.50), and 770 nm (710 nm), due to the local character of the EDX measurements. In the impossibility to perform more sophisticated measurements both for the as-deposited and the irradiated samples, we considered, for modelling purposes, EDX linear interpolated profiles as a reference and we checked the impact  of different alloy fraction and thicknesses.  \par
For our computations, we employed a pre-existing code developed by some of us \cite{lamagna2004, fisicaro2014, lombardo2017, huet2020} (see related methodological section). The time harmonic electromagnetic field, computed from Maxwell equations, mimicked the one employed experimentally, with a wavelength of 308 nm and a pulse time $\Delta t$ of 160 ns. We used a simple mono-dimensional mesh with the idea of the scheme in Figure 4a. The initial alloy fraction profile along Z was taken from the EDX measurements in Figure S2a-b. The mesh presented a total length of 4500 nm (8000 nm) for the case X=0.24 (X=0.58) and is divided into three different portions with a progressively increased grain: (i) $Si_{1-x}Ge_x$, 1300 nm (3000 nm) long with a constant alloy fraction, (ii) $Si_{1-x}Ge_x$, graded region with a length of 2500 nm (4300 nm) with variable decreasing X and (iii) $Si$, 700 nm long characterized by X=0. \par
The model delivered ED thresholds of 0.55 (0.45) $J cm^{-2}$ for X=0.24 (X=0.58) in good agreement with experimental results reported in Table S4. These quite low values, if compared to strained cases \cite{dagault2019, dagault2020_2, huet2020}, reflected the thermal properties of thick relaxed samples, where conduction is mainly ascribed to the alloy and not to the Si substrate. Alloying enhances the probability of phonon-phonon scattering events, giving rise to a U shape of the thermal conductivity, with a minimum at X=0.5. Heat conduction is therefore reduced if compared to pure Si samples with associated drop of ED thresholds \cite{Wagner_2006, maycock1967} .\par
In an attempt to evaluate whether or not melting features for laser energy densities $> 0.7 Jcm^{-2}$ met experimental findings, see Figure S4 and S5, we firstly approximated the dielectric function of liquid $Si_{1-x}Ge_x$ as a linear combination of $ \varepsilon_{l,Si}$ and $ \varepsilon_{l,Ge}$ weighted by the respective fractions $(1-X)$ and $X$. Real and imaginary dielectric function parts for the elements, were calibrated in previous work \cite{huet2020}, and we report them in Table S3. \\
Unfortunately, this approximation, thereafter called original model, delivered some inconsistency between the simulated melt depths and the experimental ones, as documented in Figure 4b-c, i.e. comparison between the magenta histograms and the black dots. More specifically, the error bar of computed MDs resulted in more than $\sim$ 20 nm for smaller EDs, while the agreement was good for higher EDs, as for $ 1.50 \, J \, cm^{-2}$. The aforementioned deviations can be explained considering the reflectivity of the melt.   As shown in Table 2, our assumption (original model) provided almost identical reflectivity values of the melt, i.e., $\sim$ 0.78 for all ranges of alloy fractions and temperatures. However, this might not be the case for X far from the two elements.  We investigated this aspect, by studying the dependency of the MD on the imaginary dielectric function part.  $\Im{(\varepsilon_l)}$ then became a hyper-parameter that linked the optical constants of $l-Si_{1-x}Ge_x$ to melt depths, enabling an extension of the previous calibration to the liquid phase. We found optimal values of $\Im (\varepsilon_l) $, reported in Table 2 (see Optimal $\Im (\varepsilon_l)$ section), for which obtained MDs overlaps with experiments. These values are associated to specific time-averaged liquid alloy fractions and temperatures, $X_l$ and $ T_l$, reported in Table 2, captured at the air-liquid interface (Z=0 mesh point). 
Results collected with this approach (Table 2) underlined an effective dependence of the optical functions on T and X. If we focus on the data for 0.24, we observe that an optimal $l-Si_{1-x}Ge_x$ reflectivity higher than $\sim 0.81$ is needed for temperature close the melting point ($T_m$), while values of $\sim$ 0.78, similar to those arising from the original model, are required for higher T. The behaviour is identical for X = 0.58. \par
Along the same lines than for solid $Si_{1-x}Ge_x$, we elaborated a semi-empirical expression for $\varepsilon_l$, accounting for the specific behaviour of the reflectivity near the melting point, $T_m$. Due to the limited availability of points, we kept the real part of the function to its original form, i.e. without a temperature dependence, and we varied only its imaginary part. In this framework we had enough degrees of freedom to reproduce the correct reflectivity required to match the experiment. \\
The $f(X,T)$ function is, at variance with the solid case, cubic on X and linear on T. The cubic dependency on the alloy fraction X was found to be necessary to effectively reproduce, at the same time, the pure elements boundaries X = 0 and X = 1, and the values of the dielectric function at X = 0.24 and X = 0.58. As a matter of fact, previous attempts with a quadratic dependency led to an inexact reproduction of the function at the points used for fitting. The temperature dependency was chosen as linear due to the limited amount of points available. However, we tested a quadratic dependency for X=0.24 and we found only a marginal impact of higher order T terms in reproducing the experiment (see Table S5). Our semi-empirical expression for $\varepsilon_l (T,X)$ is detailed by (15)-(17).
\begin{equation}
 \varepsilon_{l} (T,X)=\varepsilon_{l-Ge} \cdot f(T,X)+\varepsilon_{l-Si} \cdot [1-f(T,X)]
\end{equation}
\begin{equation}
 f(X,T)= a_2(T) \cdot X^3+a_1(T) \cdot X^2+[1-a_1 (T)-a_2 (T)]\cdot X
\end{equation}
\begin{equation}
 a_i(T)=b_i (T-T_{mGe} )+c_i
\end{equation}
where $a_i$ and $b_i$ are parameters determined by the fitting of the $\Im (\varepsilon_l)$ values in Table 2 vs $X_l$ and $T_l$. Temperature was referenced to the lowest melting point of the alloy, corresponding to l-Ge, $T_{m,Ge}$. The obtained $b_i$ and $c_i$ parameters are reported in Table 3 and a reflectivity map for the liquid phase of $Si_{1-x}Ge_x$, is shown in Figure 5.\\
The map summarizes two important findings about reflectivity of liquid $Si_{1-x}Ge_x$, (i) a maximum of R on the liquidus line appeared at X = 0.5 and (ii) R monotonously decreases with T.  Both effects are ascribed to alloying, highlighting possible alterations of the electronic structure of the liquid not experienced in the crystal, where the R value only slightly differed from the average of 0.60 (Figure 3a). This deserves further investigations. \par
We observe that, for similar ED threshold, melt depths of X=0.58 samples were always deeper than for X=0.24 because of the smaller melting point. \par
The dielectric function calibration, achieved with samples' alloy fraction and thicknesses by EDX measurements, was tested also with the nominal values of the former, as shown in Table S6. As a result of this analysis, we found different thickness of $Si_{1-x}Ge_x$ did not alter the melt depths values. Negligible variations were found changing the initial alloy fraction 0.24 to 0.20, while more, though slight, happened when moving from 0.58 to 0.50. Anyway, the error bar was lower than the one arising with the original model of dielectric constants.  We performed, for completeness, a calibration considering nominal X as initial Ge concentration (see Table S7). \par
To validate our fine-tuned model for $Si_{1-x}Ge_x$, obtained with data from relaxed samples,  we used a pre-existing data-set of strained $Si_{1-x}Ge_x$ samples published in \cite{huet2020, dagault2019, dagault2020_2} and fresh measurements performed on samples with X=0.3 (laser annealing conditions were identical to relaxed samples). These experimental data-set covers germanium contents of 0.20, 0.30 and 0.40. The samples were irradiated with a XeCl laser with pulses $\Delta t$ of 160 ns and 146 ns. \par
Modifications to our FEM model, for this validation purpose, embroiled new mesh and initial alloy profile definitions, following the idea of the scheme in Figure 6a. The mesh was characterized by 30 nm of $Si_{1-x}Ge_x$ where the alloy fraction was set as constant followed by 4470 nm of Si.  Ultimately, the graded region employed for the relaxed samples was replaced by a sharp  $Si/Si_{1-x}Ge_x$ interface of $\sim 1 \, nm$. The laser pulse was selected as 160 ns or 146 ns, depending on the samples' experimental records.
 \par
In line with experiments, we achieved higher ED thresholds (see Table S4) compared to that in relaxed samples, in the range of $1.40  -  1.55$ $Jcm^{-2}$. This is ascribed to the smaller thickness of the samples, implying thermal conduction mainly governed by the silicon buffer.

\begin{table}[b]
\caption{\label{tab:table3} Melt depths obtained for $ED > 2.00 \, J \cdot cm^{-2}$ for strained samples with a studied cut-off of the imaginary dielectric function expression. Cf. main text.}
\resizebox{\columnwidth}{!}{\begin{tabular}{ccccccc}
\toprule
$X$&$ED$&$\Delta t$&$\Im(\varepsilon_l) \, cutoff$&$R_l \, cutoff$&$MD$&$MD \, exp$\\
&$[J \cdot cm^{-2}]$&$[ns]$&&&[nm]&[nm]\\
\midrule
0.2&2.20&146&8.358&0.801&43&38\\
0.2&2.40&146&8.358&0.801&81&81\\
\bottomrule
\end{tabular}}
\end{table}

Turning to melt-depths, our computational results, reported in Figure 6b-e, confirmed, in all cases, our reflectivity fine-tuning was essential for a correct matching. Accordingly, the cyan histograms are in close agreement with the black dots (experiment), while the magenta bars (original model) are always $\sim$ 20 nm deeper.  The simulated germanium segregation profiles, drawn in Figure 6f-g along with the experimental ones, featured a good level of accuracy owing to the correct reproduction of the germanium segregation process. Some critical issues occurred as we increased the laser fluency to ED $> 2.00 \,  Jcm^{-2}$ . In this regime the liquid front exceeded the 30 nm of strained $Si_{1-x}Ge_x$ samples, entering the Si buffer region.  Liquid temperatures then reached $\sim$ 1700 K, a range where our calibration is not trained and yielded an incorrect small reflectivity value of  $\sim$ 0.75. To overcome the limitation (due to temperature expression linearity), we studied a cut-off of the semi-empirical $\Im (\varepsilon_l)$ function	yielding R = 0.80 with whom the experiment is matched  (Table 4).\\

\section{Conclusions}\label{}
In conclusion, we showed the importance of correctly reproducing the air/$Si_{1-x}Ge_x$ reflectivity of the sample, in the entire ranges of X and T, in order to realistically describe the laser melting process. We addressed the issues related to the $Si_{1-x}Ge_x$ dielectric functions calibration, where a better definition of those for the liquid phase was still missing. We fine-tuned the latter with an indirect approach, using relaxed samples' data and we found the  resulting model to yield accurate results also when strained samples are considered, achieving reliable melt depths and alloy redistribution profiles.
We described some limitation for cases where the liquid front exceeded the $Si_{1-x}Ge_x/Si$ interface, whom could anyhow be circumvented with ad-hoc studied cut-offs of the dielectric functions. \par
Another noteworthy discovery arising from our investigation relates to the distinctive reflectivity exhibited by liquid $Si_{1-x} Ge_x$. Our observations indicate that the reflectivity reaches its peak, on the liquidus line, when the germanium content is at an intermediate level, gradually diminishing as the temperature increases. This intriguing trend in reflectivity could be attributed to the liquid's metallic properties and it deserves further investigation.

\section{Acknowledgements}\label{}
We gratefully acknowledge funding from the European Union’s Horizon 2020 Research and Innovation programme under grant agreement No. 871813 MUNDFAB.








\printcredits

\bibliographystyle{cas-model2-names}

\bibliography{main.bib}

\bio{}
\endbio


\end{document}